\documentclass{nature_preprint}
\usepackage{amsmath,amssymb}
\usepackage{graphicx}
\usepackage{epstopdf}
\usepackage{caption}
\usepackage{color}

\usepackage{tabularx}
\usepackage{longtable}
\usepackage{booktabs}
\usepackage{nicefrac}

\title{Stacked topological insulator built from bismuth-based graphene sheet analogues} 

\author
{Bertold Rasche$^1$, Anna Isaeva$^1$, Michael Ruck$^{1,2}$, 
Sergey Borisenko$^3$, Volodymyr Zabolotnyy$^3$, Bernd B\"uchner$^{3,4}$, 
Klaus Koepernik$^3$, Carmine Ortix$^3$, Manuel Richter$^3$, Jeroen van den Brink$^{3,4}$\\
\\
\normalsize{$^{1}$Department of Chemistry and Food Chemistry, TU Dresden, D-01062 Dresden, Germany}\\
\normalsize{$^{2}$Max Planck Institute for Chemical Physics of Solids, D-01187 Dresden, Germany}\\
\normalsize{$^{3}$Leibniz Institute for Solid State and Materials Research, IFW Dresden, D-01069, Germany}\\
\normalsize{$^{4}$Department of Physics, TU Dresden, D-01062 Dresden, Germany}\\
\\
}

\date{}

\begin{document} 


\baselineskip18pt


\maketitle 

\begin{abstract}
Commonly materials are classified as either electrical conductors or insulators.
The theoretical discovery of topological insulators (TIs) in 2005 has fundamentally challenged this dichotomy\cite{Kane05a}. 
In a TI, spin-orbit interaction generates a non-trivial topology of the electronic band-structure dictating that its bulk is perfectly 
insulating, while its surface is fully conducting.
The first TI candidate material put forward\cite{Kane05b} 
-- graphene -- is of limited practical use since its weak spin-orbit interactions produce a band-gap\cite{Min06} of 
$\sim$0.01K. Recent reinvestigation of Bi$_2$Se$_3$ and Bi$_2$Te$_3$, however, have firmly categorized 
these materials as {\it strong} three-dimensional TI's.\cite{Zhang09,Bernevig06,Konig07,Brune11,Fu07a}
We have synthesized the first bulk material belonging to an entirely different, {\it weak},
topological  class, built from stacks of two-dimensional TI's: Bi$_{14}$Rh$_3$I$_9$.
Its Bi-Rh sheets are graphene analogs, 
but with a honeycomb net 
composed of RhBi$_8$-cubes rather than carbon atoms.
The strong bismuth-related spin-orbit interaction 
renders 
each graphene-like layer 
a TI
with a 2400K band-gap.
\end{abstract}

\pagebreak

%
Besides being new states of matter and thus of the most fundamental scientific interest, topological insulators (TIs) also hold promises for applications in for instance spintronics, based on the fact that the topological properties dictate that the metallic surface-states of TIs are spin-locked: theory predicts that the propagation direction of surface-electrons is robustly linked to their spin orientation\cite{Wu06,Fu07a,Zhang09,Brune11,Konig07,Bernevig06}. These surface-states also play a most prominent role in proposals to create Majorana fermions in microelectronic devices, the manipulation of which can be the basis for future topological quantum computing\cite{Fu08,Akhmerov09,Law09}.
%
%
In spite of this conceptual richness, a lack of equivalent advances in producing new classes of TI materials has lead materials synthesis and chemistry to  concentrate largely on further perfecting and varying upon the materials class of bismuth-based chalcogenides Bi$_2$Se$_3$ and Bi$_2$Te$_3$, which are confirmed three-dimensional (3D) TIs\cite{Xia09,Chen09,Hsieh09,Kuroda12}, and on HgTe films grown with utmost care by molecular beam epitaxy which under specific conditions form a two-dimensional (2D) TI\cite{Konig07,Brune11}.
%
%
%
We have synthesized Bi$_{14}$Rh$_3$I$_9$, which we will show to be the first member of an entirely new class of stacked 2D TIs, from a stoichiometric melt of its elements. The resulting thin black platelets are air-stable and can be easily cleaved. The synthetic procedure has been optimized taking into account the phase decomposition at the peritectic point of 441$^\circ$C and the strong dependence of the phase stability on the vapor-pressure. For a more detailed synthesis protocol we refer to the Supplementary Information (SI) .

\begin{figure}
\center{\includegraphics[width=0.5\columnwidth]{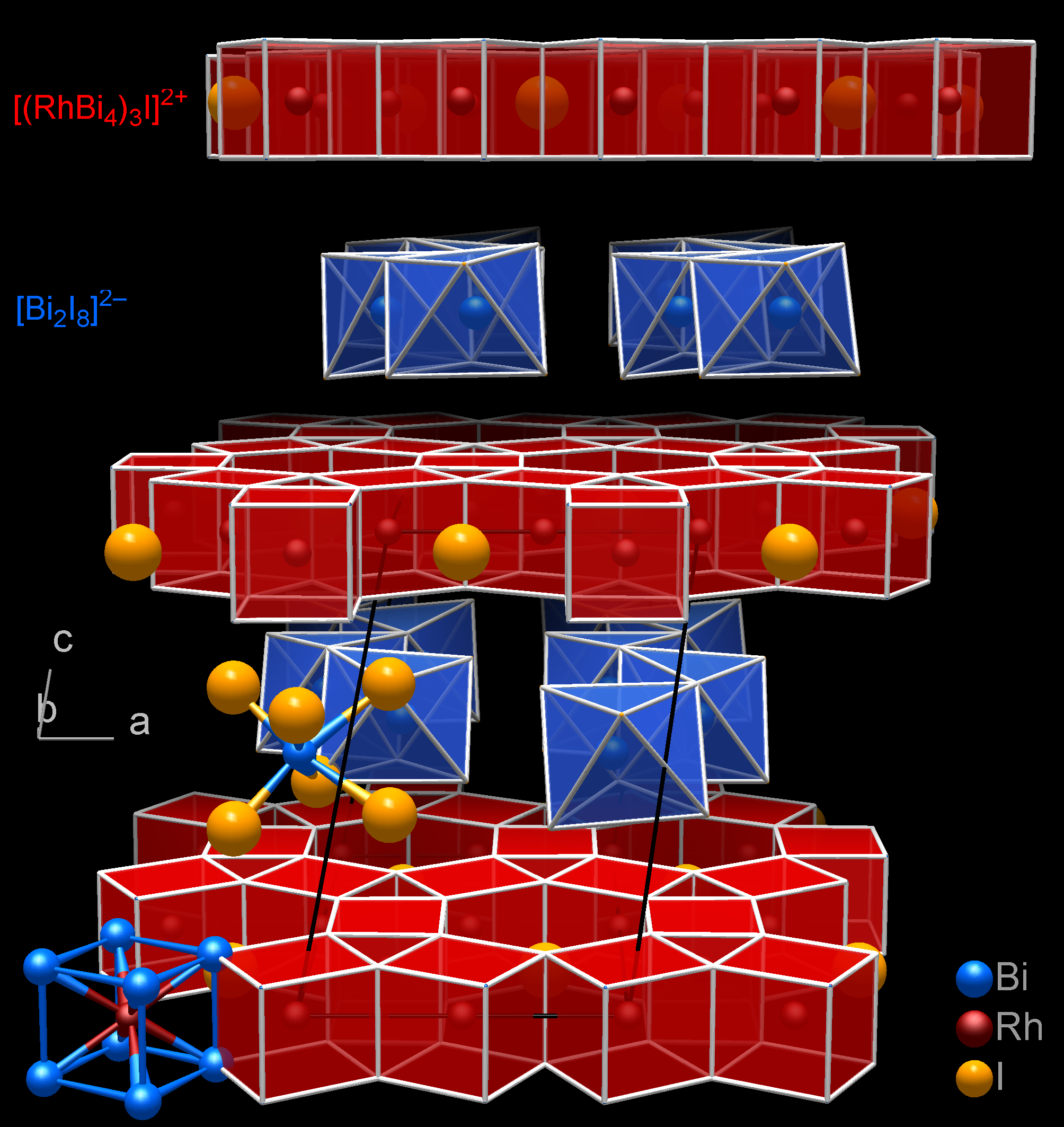}}
\center{\includegraphics[width=0.5\columnwidth]{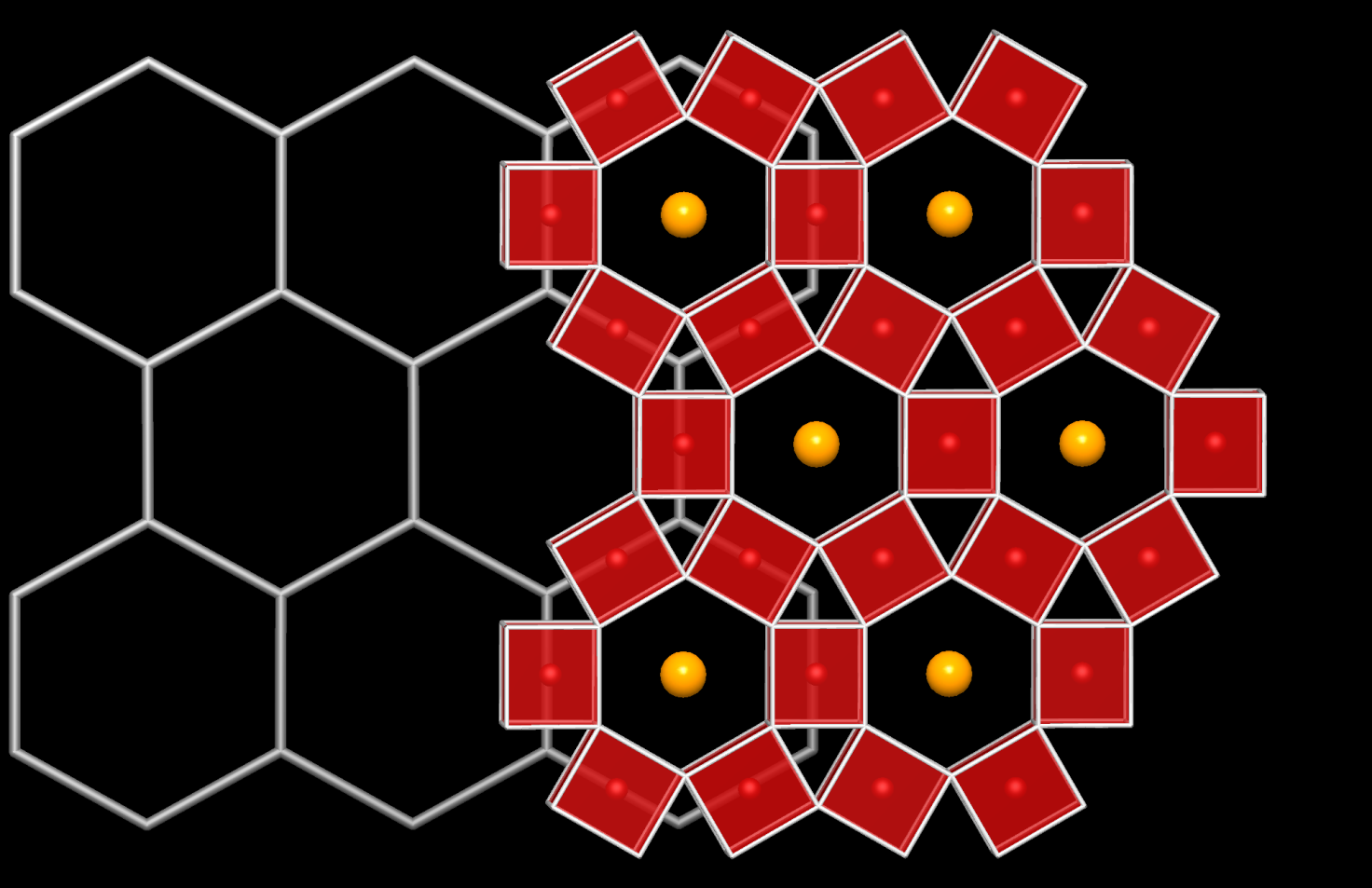}}
\caption{\baselineskip18pt
(a) Triclinic crystal structure of Bi$_{14}$Rh$_3$I$_9$. Insulating layers of [Bi$_2$I$_8$]$^{2-}$ zigzag chains separate the intermetallic layers [(Bi$_4$Rh)$_3$I]$^{2+}$ which consist of hexagonal nets of edge-sharing RhBi$_8$ cubes. (b) Honeycomb lattice of graphene scaled by a factor $\sim$3.8 overlaid with the structure of the intermetallic layer.}
\label{fig:structure}
\end{figure}

%
%
The crystal structure was determined by single-crystal X-ray diffraction and exhibits a periodic alternating stacking of 2D bismuth-rhodium networks and insulating spacers, see  Fig.~\ref{fig:structure}(a). The former, further denoted as {\it intermetallic layer}, can be understood as a decorated honeycomb network. It has the same hexagonal p6/mmm layer group symmetry as a graphene sheet, thus sharing its 24 symmetry elements. A lattice scaling of $\sim$3.8 in Fig.~\ref{fig:structure}(b) makes this structural equivalence prominent. 
Whereas the nodes of a graphene-sheet host carbon atoms, in the intermetallic layer of Bi$_{14}$Rh$_3$I$_9$ the nodes of the net are found in the centres of triangular-prismatic voids (Bi--Bi: 317.5--320.6 pm). The latter appear in the hexagonal arrangement of 
bismuth cubes (Bi--Bi: 317.5--345.3 pm) which are centred by rhodium atoms  (Rh--Bi: 282.0--284.9 pm). This arrangement can also be seen as a kagome-type net with the rhodium atoms of edge-sharing cubes at the nodes of the net. Consequently, the RhBi$_8$-cubes define the hexagonal-prismatic voids which are filled with iodide anions (Bi$\cdots$I: 376.2--382.1 pm). As a result, an overall composition of (RhBi$_4$)$_3$I can be assigned to the intermetallic layer. 
Chemical-bonding analysis reveals strongly localized, covalent Bi--Rh bonds in the cubes and three-centred bismuth interactions in the bases of the triangular-prismatic voids, all together establishing a quasi-2D bimetallic network (see SI for details). 
The spacer layer consists of Bi-I zigzag chains with distorted octahedral coordination of bismuth(III) cations by iodide anions.
Bismuth-iodine bonding between the intermetallic layer and the spacer (distances 376.4-455.6 pm, mean value 401.9 pm) point at weak interactions of charged layers and render the whole sandwiched structure of Bi$_{14}$Rh$_3$I$_9$ = [(RhBi$_4$)$_3$I]$^{2+}$[Bi$_2$I$_8$]$^{2-}$ as salt-like in stacking direction. The alternate stacking of the highly-symmetric intermetallic and the low-symmetric spacer layer results in the reduction of the overall crystal symmetry to the triclinic space group $P\bar{1}$.

\begin{figure}
\center{
\includegraphics[width=0.95\columnwidth]{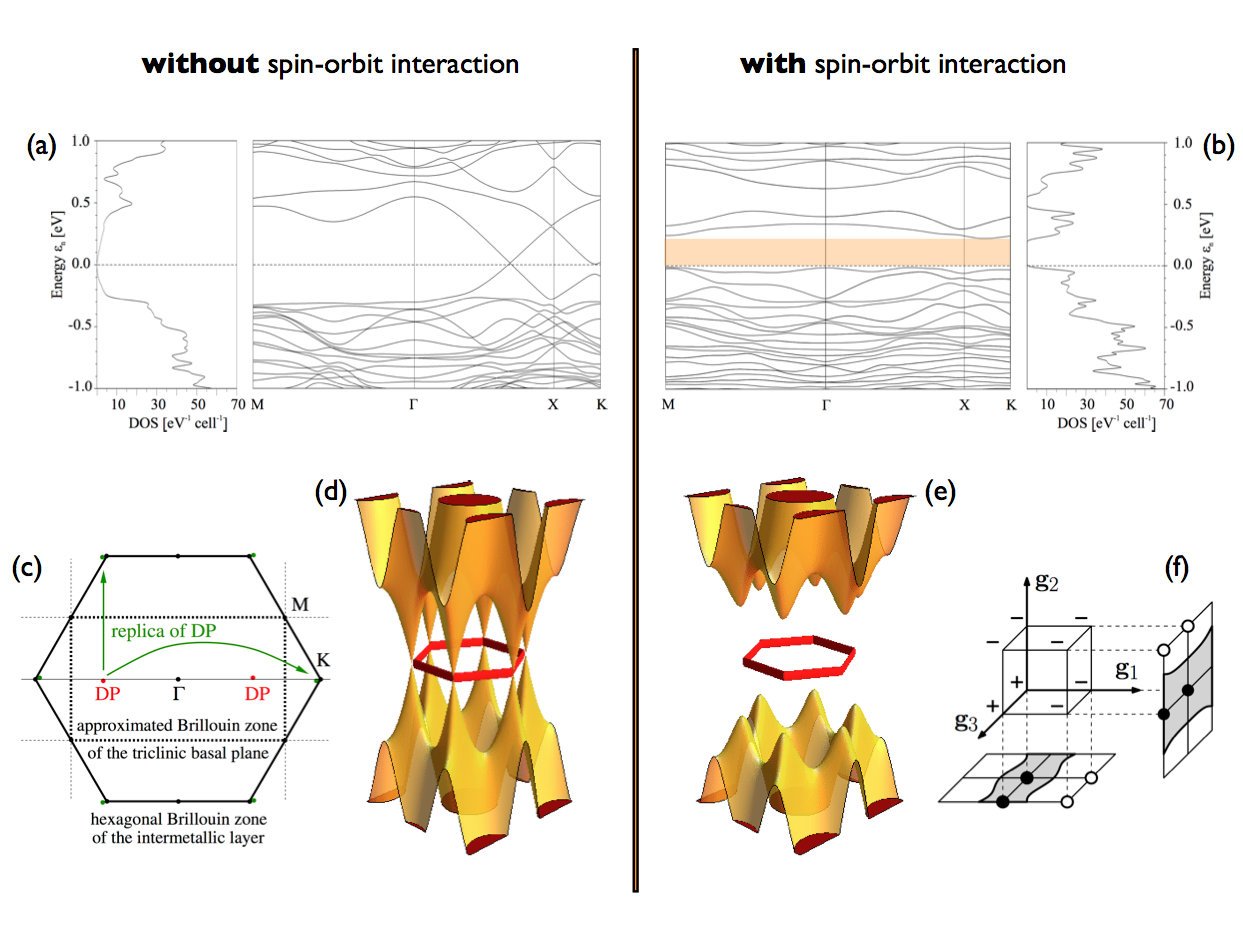}
}
\caption{\baselineskip18pt
(a) Scalar relativistic band-structure of triclinic Bi$_{14}$Rh$_3$I$_9$. On the line $\Gamma$X a Dirac cone is present. (b) Fully relativistic band-structure where spin-orbit interaction opens up a gap of $\sim$210 meV (shaded area).  (c) Positions of Dirac points (DP) in the basal-plane of the triclinic BZ (dashed lines) and the hexagonal BZ (full lines). DPs are on the line $\Gamma$K in the triclinic BZ, corresponding to a position close to the K points of the unfolded hexagonal BZ. (d) Illustration of the Dirac cones in the situation without spin-orbit interaction and  (e) the topological gap induced by spin-orbit interaction. (f) Parity eigenvalues at the 8 time reversal invariant points in the 3D BZ, where $g_3$ corresponds to the direction perpendicular to the Bi-Rh planes. The parity eigenvalues lead to the topological invariants $\nu_0 ; (\nu_1,\nu_2,\nu_3) = 0; (0,0,1)$. Projected parities on the planes perpendicular to $g_1$ and $g_2$ illustrate the presence of topological edge-states on these surfaces. 
}
\label{fig:bands}
\end{figure}

%
%
The fact that the weakly coupled intermetallic layers have the same structural symmetry as graphene-sheets suggests similarities in electronic structure between graphene and Bi$_{14}$Rh$_3$I$_9$.
Indeed a scalar-relativistic band-structure calculation for Bi$_{14}$Rh$_3$I$_9$, where the spin-orbit coupling (SOC) is effectively switched off, (for details see SI), reveals the presence of two Dirac cones in the triclinic Brillouin-zone (BZ), which are situated at the Fermi-level, see Fig.~\ref{fig:bands}a. 
Unfolding the triclinic zone to a hexagonal one, we observe that the two inequivalent Dirac cones appear very close to the $K$ and 
$K'$ point at the edge of the hexagonal BZ, precisely where they are in graphene (Fig.~\ref{fig:bands}c).  
The very weak hopping of electrons between the layers causes the Dirac cones to pick up a minor dispersion perpendicular to the plane, rendering the calculated band-structure quasi-2D.
%
%
%

However, for a proper understanding of the electronic band-structure one needs go beyond calculations in scalar relativistic approximation.
Whereas the relativistic SOC in for instance graphene is very weak -- it is expected to open a gap on an energy scale\cite{Min06} of 0.01 K -- bismuth is well-known for its strong SOC which can drive and stabilize topologically non-trivial electronic states. As we are thus dealing with graphene-like Dirac cones in the presence of strong SOC, the mechanism proposed by Kane and Mele\cite{Kane05b} to stabilize a topologically non-trivial quantum spin Hall state will be in action full-force in the intermetallic layers of Bi$_{14}$Rh$_3$I$_9$. Indeed, a fully-relativistic band-structure calculation reveals a gapping out the Dirac cones, resulting in a calculated band-gap of 210 meV, corresponding to 2400 K, see Fig.~\ref{fig:bands}b. 

To confirm the topological nature of the resulting insulating state we have implemented the direct calculation of the four topological $Z_2$ invariants\cite{Fu07a,Fu07b,Moore07,Roy09} $\nu_0;(\nu_1,\nu_2,\nu_3)$ in the FPLO band-structure code\cite{Koepernik99}  (see SI for details). This calculation is based on an analysis of the wavefunction parity eigenvalues at the eight time-reversal invariant points of the band-structure, as illustrated in Fig.~\ref{fig:bands}f. Because of the stacking of the quasi-2D intermetallic planes in which SOC has gapped out the Dirac cones, one expects Bi$_{14}$Rh$_3$I$_9$ to be a weak TI and indeed we find  $\nu_0=0$. The other topological invariants are calculated to be $(\nu_1,\nu_2,\nu_3)=(0,0,1)$.  The fact that $\nu_0=0$ and $\nu_3=1$ proves that Bi$_{14}$Rh$_3$I$_9$ is a weak TI -- the first synthesized material in this topological class -- and  confirms that the intermetallic planes form sheets of Quantum Spin Hall states that are stacked along the $c$-axis.
It implies that in the Altland-Zirnbauer classification\cite{Altland97} Bi$_{14}$Rh$_3$I$_9$ belongs to the symplectic (AII) topology class.
On surfaces perpendicular to $(0,0,1)$  -- and therefore parallel to the intermetallic planes, which are the natural cleaving planes of the material -- topological surface-states will thus be absent. At any other surface an even number of Dirac cones appears, each having a strongly anisotropic group velocity due to the quasi-2D nature of the bulk band-structure. The metallicity of these surface-states is stable against disorder, which, as for strong TIs in the symplectic class, does not act as a source of localization\cite{Mong12,Ringel12}.  

\begin{figure}
\center{\includegraphics[width=0.75\columnwidth]{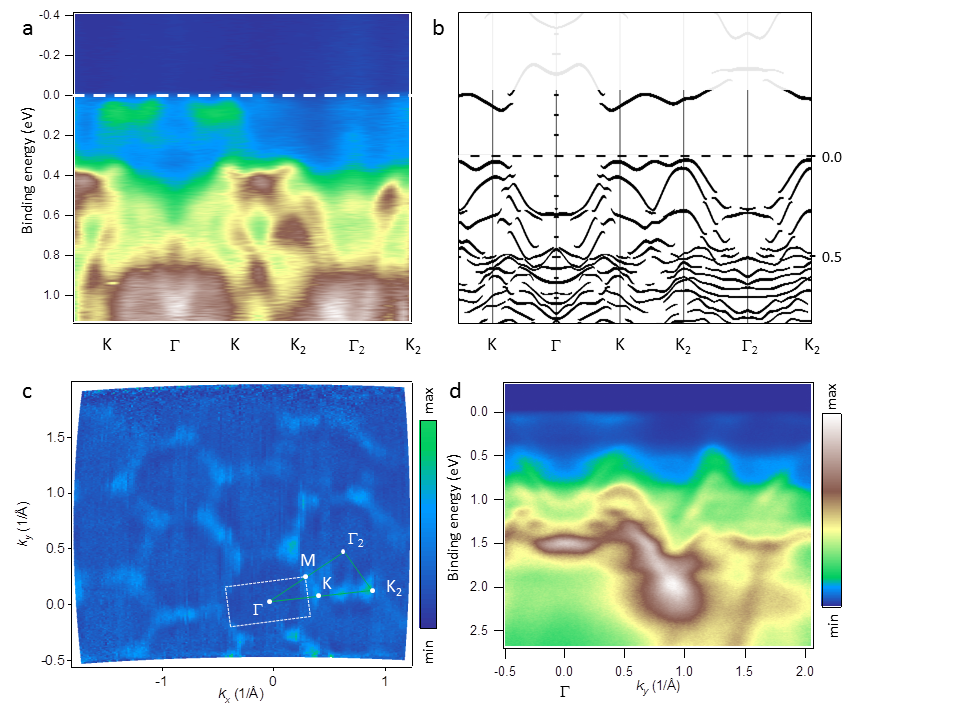}}
\vspace{2cm}
\caption{\baselineskip18pt
Band-structure as measured by angle-resolved photo-emission\cite{Borisenko12} compared to the calculated one. A small amount of electron doping brings the Fermi-level into the conduction band, so that the topological band-gap becomes clearly visible: (a) Momentum-energy intensity plot along the high-symmetry $\Gamma$K direction of the unfolded Brillouin zone. (b) Unfolded fully relativistic band-structure along the same direction. (c) Momentum distribution at 0.4 eV binding energy integrated within 50 meV window. White dashed lines show the trigonal, surface projected BZ. (d) Typical momentum-energy distribution taken along the cut which corresponds to $k_x=0$ in panel (c).
}
\label{fig:arpes}
\end{figure}

%
%
The band-structure obtained from the electronic structure calculations can be tested experimentally by Angle Resolved Photoemission (ARPES), which in particular can provide the experimental value of the electronic band-gap, if by slight electron doping the Fermi-level of the material is pushed into the conduction band. The ARPES spectra in Fig.~\ref{fig:arpes}a,d, show that indeed the material is n-doped, which is consistent with a slight iodine-deficiency related to its vapor-pressure or alternative causes for charge redistribution at the surface probed by ARPES. It offers a unique opportunity to check the requisites of the topological scenario experimentally, in a full analogy with for instance the well-known case\cite{Xia09} of Bi$_2$Se$_3$. 
 
We compare the ARPES results with the fully relativistic band-structure calculations in the unfolded BZ selecting  the $\Gamma$K high-symmetry direction (Fig.~\ref{fig:arpes}a,b, for other cuts see SI). This is a direction where the Dirac cones are found in scalar-relativistic approximation. The agreement between experimental data and fully-relativistic calculations is remarkable. First of all, the observed gap, as determined considering the distance between the features of the integrated (over three BZ) spectral weight, is $\sim$270 meV, which is consistent with the relativistic band-structure. Also the number and behavior of the dispersing features clearly seen in ARPES intensity plots are essentially captured by the bands projected to the hexagonal BZ. The prominent examples are the inequivalence of $\Gamma$ and $\Gamma_2$ points and structures between K and K$_2$, both also present in the calculated band-structure. 

This agreement confirms experimentally  the correctness of the calculated band-structure in presence of spin-orbit coupling -- the bands for which we have calculated the topological invariants to be $\nu_0 ; (\nu_1,\nu_2,\nu_3) = 0; (0,0,1)$. This topologically non-trivial state we have understood in terms of the graphene-like structure of the intermetallic Bi-Rh planes, for which two Dirac cones are present at the Fermi-level in a calculation without spin-orbit coupling. Experimentally the spin-orbit coupling is of course unavoidable and switching it on in the calculation gaps out the Dirac cones and generates the weak topological $0; (0,0,1)$ state. As a consequence spin-locked topological surface states will be present at any crystal face that is {\it not} parallel to the $(001)$ plane. Such faces, however, do not correspond to natural cuts of the crystal and the associated surface roughness has prevented us so far from observing these spin-polarized surface states by ARPES on, for instance, $(100)$ surfaces. In principle (spin-polarized) scanning tunneling microscopy and spectroscopy should be able to directly probe these states at step-edges of the natural $(001)$ cleaving plane.

Finally, the momentum distribution of  ARPES intensity at 400 meV binding energy, just below the calculated Fermi level of Bi$_{14}$Rh$_3$I$_9$, clearly shows a hexagonal pattern (Fig.~\ref{fig:arpes}c), as is expected from the band-structure calculations where the top of the valence band is formed by a dispersive feature that is relatively flat between the K points. The hexagonal pattern justifies the unfolding of the BZ carried out to facilitate the comparison with ARPES. At the same time its slight irregularity reflects the over-all triclinic crystal symmetry. 

The hexagons observed in ARPES emphasize the structural and electronic similarities of graphene and Bi$_{14}$Rh$_3$I$_9$, which again it shares with the chemically and structurally closely related compound\cite{Ruck97,Ruck01} Bi$_{13}$Pt$_3$I$_7$, for instance. The compelling difference with graphene is that the large spin-orbit coupling drives Bi$_{14}$Rh$_3$I$_9$ electronically into a topologically insulating state, corresponding to a 3D stack of 2D quantum spin Hall states, which is very different from the strong TI states observed so far in bulk materials such as Bi$_2$Se$_3$/Bi$_2$Te$_3$ and which is predicted to leave marked signatures on electronic transport through its spin-polarized, stacked quasi-1D topological surface-states\cite{Mong12,Ringel12}.

\end{document}